\documentclass [a4paper,11pt]{article}
\usepackage[margin=2cm, letterpaper]{geometry}
\usepackage{indentfirst}
\usepackage{hhline}
\usepackage{wrapfig}
\usepackage{graphicx,amsmath,amssymb}
\usepackage{array,longtable,subfigure,tabularx}
\usepackage[small]{caption2}
\usepackage{subfigure}
\usepackage{color}
\usepackage[colorlinks, urlcolor=blue, linkcolor=blue, citecolor=blue]{hyperref}
\hypersetup{urlcolor=blue, linkcolor=blue, citecolor=blue, colorlinks=true}
\renewcommand{\captionlabeldelim}{.}
\begin{document}

\begin{center}

%
%
%
%
%

{\bfseries Charge transport in Ho$_x$Lu$_{1-x}$B$_{12}$: Separating
Positive and Negative Magnetoresistance in Metals with Magnetic
Ions }

\vspace{1cm}

\bigskip \bfseries  N.E. Sluchanko$\!\!\!\phantom{x}^{a,b,\star}$, A.L. Khoroshilov$\!\!\!\phantom{x}^{a,b}$, M. A. Anisimov$\!\!\!\phantom{x}^a$, A.N. Azarevich $\!\!\!\phantom{x}^{a,b}$, A.V. Bogach$\!\!\!\phantom{x}^{a}$, V.V. Glushkov$\!\!\!\phantom{x}^{a,b}$, S.V. Demishev$\!\!\!\phantom{x}^{a,b}$, V.N.Krasnorussky$\!\!\!\phantom{x}^{a}$, V.V. Voronov$\!\!\!\phantom{x}^{a}$, N.Yu. Shitsevalova$\!\!\!\phantom{x}^{c}$, V.B. Filippov$\!\!\!\phantom{x}^{c}$, A.V. Levchenko$\!\!\!\phantom{x}^{c}$, G. Pristas$\!\!\!\phantom{x}^{d}$, S. Gabani$\!\!\!\phantom{x}^{d}$, K. Flachbart$\!\!\!\phantom{x}^{d}$ \vspace{0.5cm}\mdseries
\end{center}

\begin{center}

\slshape{\mdseries{$\phantom{x}^a$ \textit{ A. M. Prokhorov General Physics Institute,
 Russian Academy of Sciences, 38 Vavilov str., Moscow, 119991 Russia}

 $\phantom{x}^b$ \textit{ Moscow Institute of
Physics and Thechnology, Institutskii per. 9, Dolgoprudnyi, Moscow
Region 141700, Russia}

 $\phantom{x}^c$ \textit{I. N. Frantsevich Institute for Problems of Materials Science, National Academy of Sciences of Ukraine, 3 Krzhizhanovskii str., Kiev, 03680 Ukraine}

 $\phantom{x}^d$ \textit{ Institute of Experimental Physics of Slovak Academy of Sciences, 47 Watsonova str., 040 01 Ko\v{s}ice, Slovak Republic}
 }}

 e-mail: nes@lt.gpi.ru

\end{center}
\begin{abstract}
The magnetoresistance (MR) $\Delta \rho/\rho$ of cage-glass
compound Ho$_x$Lu$_{1-x}$B$_{12}$ with various concentration of
magnetic holmium ions ($x$$\leq$0.5) has been studied in detail
concurrently with magnetization $M(T)$ and Hall effect
investigations  on high quality single crystals at temperatures
1.9-120 K and in magnetic field up to 80 kOe. The undertaken
analysis of $\Delta\rho/\rho$ allows us to conclude that the
large negative magnetoresistance (nMR) observed in vicinity of
Neel temperature is caused by scattering of charge carriers on
magnetic clusters of Ho$^{3+}$ ions, and that these nanosize
regions with AF exchange inside may be considered as short range
order AF domains. It was shown that the Yosida relation $-\Delta
\rho/\rho$$\sim$$M^2$ provides an adequate description of the nMR
effect for the case of Langevin type behavior of magnetization.
Moreover, a reduction of Ho-ion effective magnetic moments in the
range 3-9$\mu_B$ was found to develop both with temperature
lowering and under the increase of holmium content. A
phenomenological description of the large positive quadratic
contribution $\Delta \rho/\rho$$\sim$$\mu_D^2 H^2$ which dominates in
Ho$_x$Lu$_{1-x}$B$_{12}$ in the intermediate temperature range
20-120 K allows to estimate the drift mobility exponential changes
$\mu_D$$\sim$$T^{-\alpha}$ with $\alpha$=1.3-1.6 depending on Ho
concentration. An even more comprehensive behavior of
magnetoresistance has been found in the AF state of
Ho$_x$Lu$_{1-x}$B$_{12}$ where an additional linear positive
component was observed and attributed to charge carriers
scattering on the spin density wave (SDW). High precision
measurements of $\Delta\rho/\rho$$=$$f(H,T)$ have allowed us also to
reconstruct the magnetic \textit{H-T} phase diagram of
Ho$_{0.5}$Lu$_{0.5}$B$_{12}$ and to resolve its magnetic structure
as a superposition of 4\textit{f} (based on localized moments) and
\textit{5d} (based on SDW) components.
\end{abstract}

{\bfseries Keywords:} boron compounds, dodecaborides, magnetoresistance, magnetic clusters, antiferromagnet, spin density wave, local susceptibility

\section*{I. Introduction}

Magnetoresistance (MR) as a property of a material to change the
value of its resistivity in external magnetic field was discovered
by Lord Kelvin in 1856 \cite{1}, but the mechanisms which are
responsible both for negative and positive MR effects in various
materials are still a subject of debate \cite{2}-\cite{10}. Over
the past two decades a number of materials with large MR, such as
organic semiconductors \cite{10, 11}, pregraphitic carbon
nanofibers, hydrogenated and fluorinated graphene
\cite{12}-\cite{14}, amorphous Si doped with magnetic rare earth
ions \cite{15} and bulk germanium doped by multiply charged
impurities \cite{16}, SnO$_2$ (Ref.\cite{17}), silver
chalcogenides \cite{4, 18, 19} zero-band-gap Hg$_{1-x}$Cd$_x$Te
(Ref. \cite{20}), frustrated metallic ferromagnets \cite{21} etc.,
which are characterized by extreme field sensitivity and/or large
values of MR, have been studied in detail, because of their
potential for technological applications as e.g. magnetic sensors
and/or magnetoresistive reading heads in magnetic recording
\cite{22}. Special attention has been paid also to several types
of compounds with magnetic \textit{d}- or \textit{f}-ions having
"colossal" negative magnetoresistance (CMR) as e.g. manganites
\cite{23, 24} and cobaltites \cite{25}, double perovskites
\cite{26}, europium-based hexaborides \cite{27}, manganese oxide
pyrochlores \cite{3, 28}, Cr-based chalcogenide spinels \cite{29,
30}, chromium dioxides \cite{31}, GdSi (Ref.\cite{32}), MnSi
(Ref.\cite{33}), CeB$_6$ and CeAl$_2$ (Refs.\cite{34, 35}), Zintl
compound Eu$_{14}$MnBi$_{11}$ (Ref. \cite{36}) etc., where the MR
reaches its largest value near ferro- or antiferromagnetic phase
transitions and is quite temperature dependent in this region.
Some of the aforementioned compounds are half-metals (metallic for
one spin orientation of the carriers while insulators for the
other orientation), others are degenerate magnetic semiconductors
or magnetic metals. At least during the last decade it became
evident that colossal magnetoresistance is not exclusive for
manganites and debates are emerging that maybe there is some
common explanation for these materials beyond (independent on)
various structures and/or interactions which characterize each of
them. In fact, various types of imperfections (substitution
disorder, vacancies and other lattice defects, electronic,
magnetic and structural inhomogeneities, non-stoichiometry, phase
separation, etc.) can be found in these compounds
\cite{11}-\cite{36}. In particular, in the case of manganites both
(\textit{i}) nanometer scale coexisting magnetic clusters and
(\textit{ii}) disorder-induced phase separation with percolative
characteristics between equal-density phases, together with
short-range polaron formation are the main factors for the
dramatic inhomogeneity resulting into a strong influence of
external magnetic field and into appearance of the CMR effect
\cite{24}. As a result, it is commonly believed at present that a
coexistence of various ordered and disordered phases plays the key
role in the CMR effect. Moreover, it is argued that the colossal
negative magnetoresistance in compounds with magnetic ions is in
fact a Griffits phase singularity arising in thermodynamic
properties at $T_M^{rand}$$\leq$$T$$\leq$$T_G$, i.e. between the random
transition temperature $T_M^{rand}$ and the "pure" transition
temperature $T_G$ (Ref. \cite{37}), and that the vicinity to
percolation threshold is the key factor to reach the CMR. In
addition, the suppression of short-range static and dynamic
polaron correlations in magnetic field is considered as another
important component to provide the negative MR effect which
dominates in magnetic materials just above $T_M$.

\begin{figure*}[h]
\begin{center}
\begin{minipage}[h]{0.79\linewidth}
\center{\includegraphics[width=0.79\linewidth]{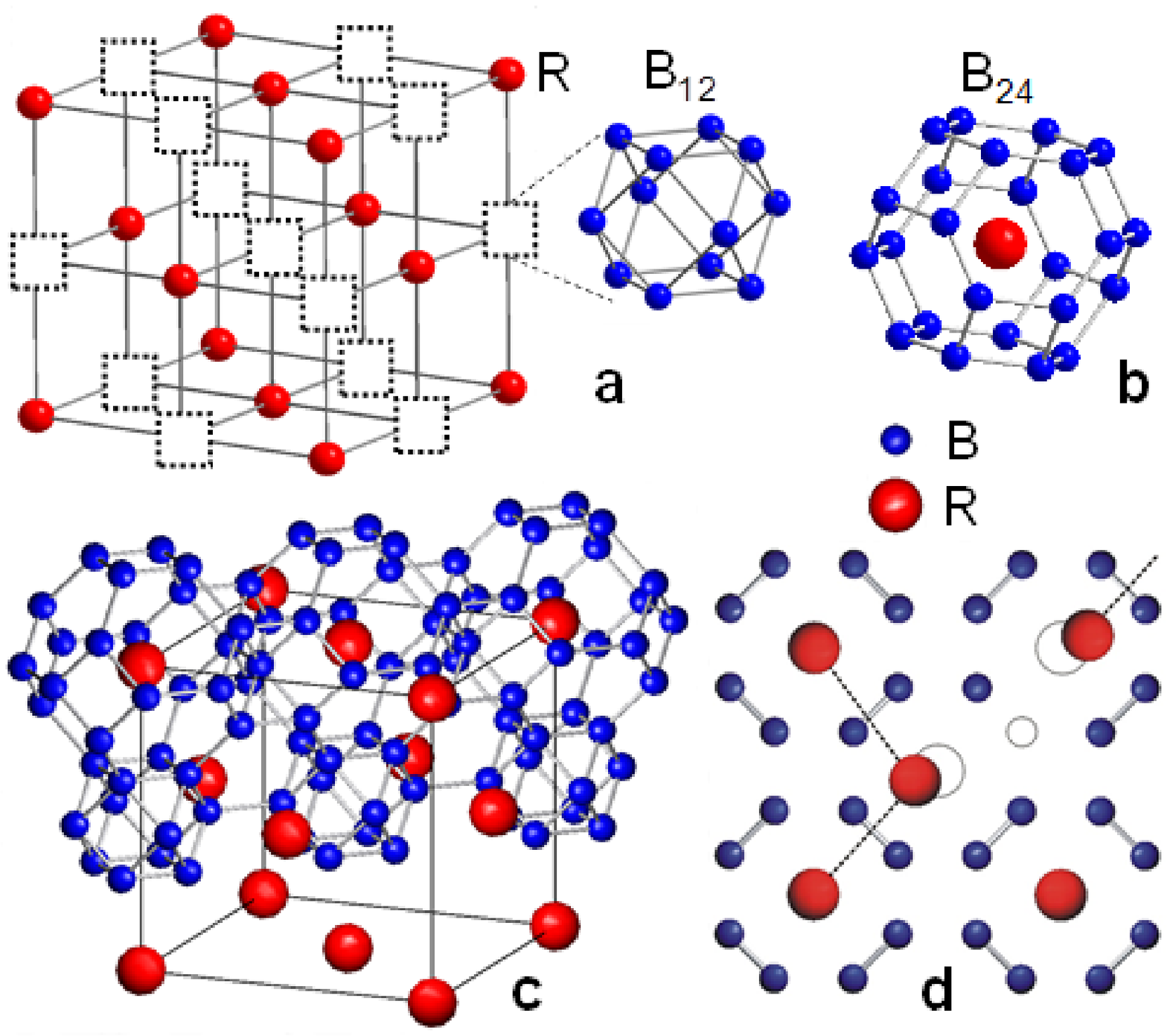}}
\end{minipage}
\end{center}
\begin{minipage}[h]{0.5\linewidth}
\center{\includegraphics[width=0.45\linewidth]{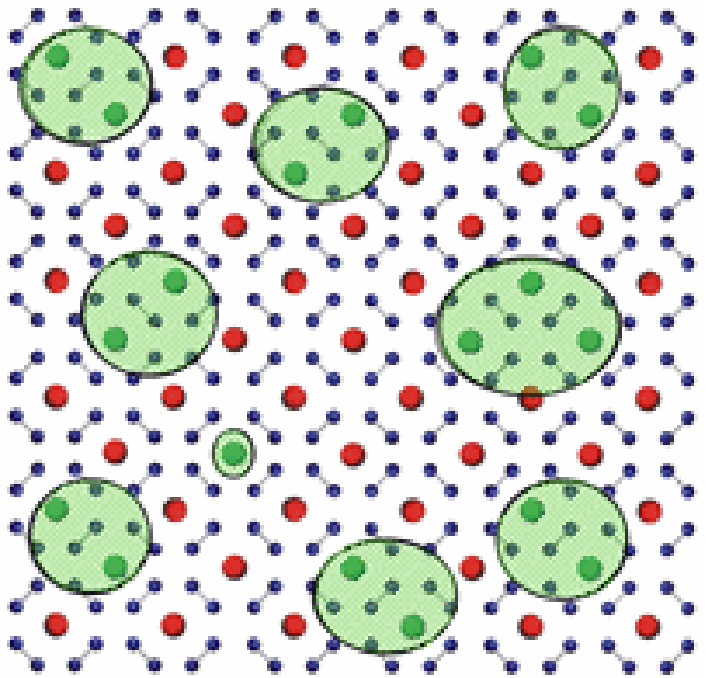}\\\textbf{e}}
\end{minipage}
\begin{minipage}[h]{0.5\linewidth}
\center{\includegraphics[width=0.45\linewidth]{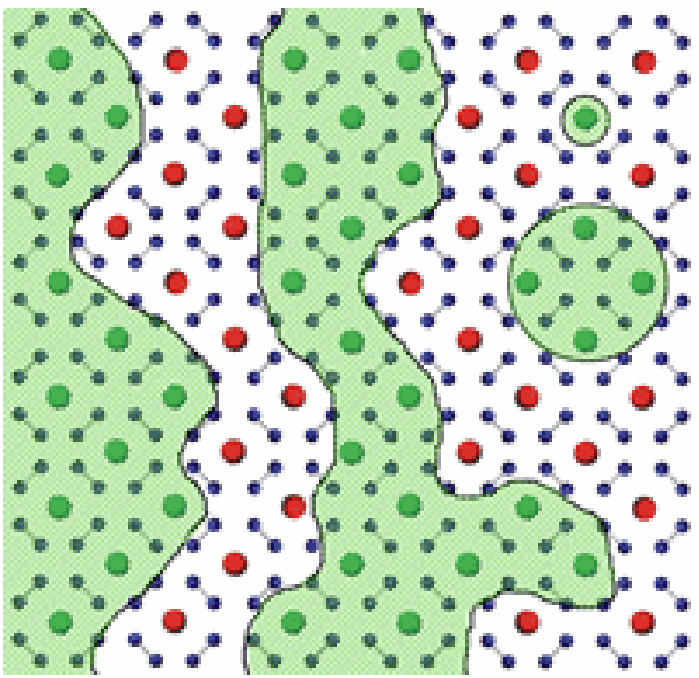}\\\textbf{f}}
\end{minipage}
\caption{(a) Crystal structure of Ho$_x$Lu$_{1-x}$B$_{12}$ compounds. The NaCl-type unit cell is built from R$^{3+}$ ions and B$_{12}$ cubooctahedrons. (b) The first coordination sphere of R$^{3+}$ is arranged as a truncated octahedron B$_{24}$. The arrangement of R and B atoms along the direction
$\langle 110 \rangle$ and in the (110) section is presented in (c) and (d), respectively. For clarity,
$B_{12}$ and $B_{24}$ clusters are shown in (c) only along the upper face diagonal of the lattice. A lattice defect (boron vacancy) is shown (small open circle) in (110) section (d). Broken \textit{R-B} bonds in the vicinity of boron vacancy cause displacements of the nearest R$^{3+}$ ions away from the defect by 0.4${\mathrm{\AA}}$.
As a result, random displacements of the R$^{3+}$ ions, R$^{3+}$ dimers and other small-size rare earth clusters (shown in (d)) appear in RB$_{12}$ matrix. (e) Magnetic clusters of holmium ions in Ho$_x$LuB$_{12}$ solid solutions in the dilute limit $x$$\leq$0.1. (f) Formation of infinite magnetic clusters of holmium ions when holmium concentration $x$(Ho) exceeds the value of the percolation threshold $x_C$ (see the text).
}\label{FigX1}
\end{figure*}

To shed more light on the origin of negative magnetoresistance
observed in strongly correlated electron systems at the vicinity
of magnetic phase transitions it is useful to investigate model
compounds with a quite simple crystalline and magnetic structure
where both different types of disorder and a dispersion of size
and number of magnetic clusters can be formed and controlled in
the vicinity of magnetic phase transition. As promising materials
for the study of negative MR effect we have chosen the
\textit{fcc} metallic substitutional solid solutions
Ho$_x$Lu$_{1-x}$B$_{12}$ with Ho magnetic ions embedded in a rigid
covalent boron cage of the dodecaboride lattice. Comprehensive
investigations of high-quality single crystals of LuB$_{12}$ with
various boron isotope compositions allowed recently to find a new
disordered "cage-glass" phase at liquid nitrogen temperatures
\cite{38}-\cite{40}. It was shown \cite{39, 40} that the
combination of loosely bound states of rare-earth ions in the
rigid boron sub-lattice of RB$_{12}$ compounds (Fig.\hyperref[FigX1]{1a-c})
together with randomly arranged boron vacancies (with a
concentration $\sim$1-3$\%$) (Fig.\hyperref[FigX1]{1d}) leads to a development of
lattice instability at intermediate temperatures. As a result, in
the range $T$$<$$T^{\star}$$\sim$60 K metallic R$^{3+}$-ions become
frozen in randomly distributed off-center positions inside
truncated $B_{24}$ octahedrons (Fig.\hyperref[FigX1]{1b-d}). In case of solid
solutions Ho$_x$Lu$_{1-x}$B$_{12}$ with magnetic rare earth ions,
there is also Lu to Ho substitutional disorder which interferes
with the random displacements (static disorder) of R-sites in the
metallic cage glass phase.

In previous magnetoresistance measurements of the non-magnetic
reference compound LuB$_{12}$ and of the antiferromagnetic (AF)
HoB$_{12}$ in AF and paramagnetic (P) states a large negative MR
effect (of about 30$\%$ in magnetic field $H$$\sim$80 kOe) was
observed in HoB$_{12}$ in the vicinity of AF-P phase transition
\cite{41}. Taking into account that holmium to lutetium
substitution in Ho$_x$Lu$_{1-x}$B$_{12}$ system is accompanied
with Neel temperature lowering from $T_N$$\approx$7.4 K for $x$=1 to
$T_N$$\approx$1.9 K for $x$=0.3 \cite{42, 43} and that at least for
$x$$\leq$0.1 a  paramagnetic ground state is detected in these
dodecaborides, it becomes possible to investigate the emergence of
the negative MR effect in absence of AF long range order in
diluted magnetic compounds with small nanosize magnetic clusters
of Ho-ions embedded in the boron matrix (Fig.\hyperref[FigX1]{1e}). Moreover, an
infinite magnetic cluster of holmium ions is expected to appear
when the $x$(Ho) concentration exceeds the percolation threshold
value $x_C$ in the \textit{fcc} lattice (Fig.\hyperref[FigX1]{1f}), which is
accompanied by AF ground state formation in
Ho$_x$Lu$_{1-x}$B$_{12}$ in the range 0.2$<$$x$$<$0.3. Thus, the role
of different size magnetic clusters during the emergence of large
negative MR effect may be investigated in details in vicinity of
AF-P phase boundary in these cage-glass metals.

The aim of this work was to perform a comparative study of
transverse magnetoresistance both for diluted magnetic ($x$=0.01,
0.04, 0.1, 0.15 and 0.19) and more concentrated antiferromagnetic
($x$=0.23, 0.3 and $x$=0.5, Ref. \cite{42})
Ho$_x$Lu$_{1-x}$B$_{12}$ solid solutions in the temperature range
1.9-100 K and in magnetic fields up to 80 kOe. In parallel, we
have performed also Hall effect measurements at $H$=80 kOe for $x$=0.1 and $x$=0.5 to
compare the drift and Hall mobility of charge carriers, and to
clarify the origin of the negative and positive MR components.
Additionally, to provide a link between the negative MR and
magnetic properties of these model dodecaborides, we have
investigated the magnetic susceptibility of
Ho$_x$Lu$_{1-x}$B$_{12}$ in a wide temperature range 2-300 K at
small magnetic field $H$$\leq$5 kOe.

\section*{II. EXPERIMENTAL DETAILS  }

In the present study, detailed investigations of resistivity,
transverse magnetoresistance and Hall effect of high-quality
single crystalline samples of Ho$_x$Lu$_{1-x}$B$_{12}$ solid
solutions with $x$=0.01, 0.04, 0.1, 0.15, 0.19, 0.23, 0.27, 0.3 and 0.5
were performed in a wide temperature range (1.9-100 K) and in
magnetic fields of up to 80 kOe (\textbf{\textit{H}}$\parallel$$\langle
001\rangle$). Resistivity and Hall resistance were measured by
standard DC five probe technique with the orientation of measuring
current \textbf{\textit{I}}$\parallel$$\langle110\rangle$. The magnetic
susceptibility was measured by a commercial SQUID-magnetometer
MPMS-5 (Quantum Design). The single crystals used for measurements
were grown by vertical crucible-free inductive floating zone
melting with multiple re-melting in an inert gas atmosphere on a
setup described in detail in \cite{44}. The high accuracy
0.01-0.02 K of temperature control of the sample holder, which was
required to perform numerical differentiation of the experimental
curves of magnetoresistance $\Delta\rho/\rho$$=$$f(H, T_0)$ with
respect to magnetic field, was achieved with the help of
commercial temperature controller TC 1.5/300 (Cryotel Ltd.) in
combination with a thermometer CERNOX 1050 (Lake Shore
Cryotronics, Inc.).

\section*{III.  EXPERIMENTAL RESULTS}
\subsection*{\emph{IIIa. Resistivity of Ho$_x$Lu$_{1-x}$B$_{12}$.}}

\begin{figure*}
\begin{center}
\includegraphics[width = 10cm]{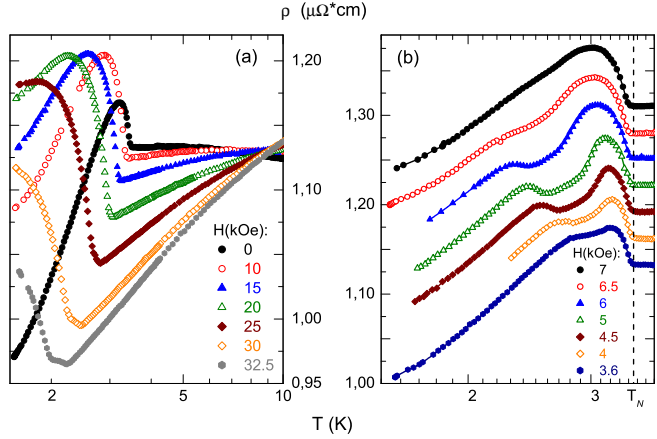}
   \caption{(a-b) The temperature dependences of electrical resistivity $\rho(T)$ of Ho$_{0.5}$Lu$_{0.5}$B$_{12}$ compound recorded in magnetic fields $H$$\leq$40 kOe. On panel (b) the $\rho(T)$ curves are shifted by a constant value 0.3 $\mu\Omega\cdot$cm for convenience.}\label{FigX2}
   \end{center}
\end{figure*}

Figure \hyperref[FigX2]{2}  shows the temperature dependences of electrical
resistivity $\rho(T)$ of Ho$_{0.5}$Lu$_{0.5}$B$_{12}$ crystal
measured in various magnetic fields below 40 kOe in a wider
vicinity of the antiferromagnetic - paramagnetic (AF-P) transition.
The found Neel temperature $T_N$$\approx$3.45 K (Fig.\hyperref[FigX2]{2}) coincides
with a good accuracy with results received in \cite{42, 43}. As
can be seen from Fig.\hyperref[FigX2]{2}, below $T_N$ the resistivity rises to a
maximum upon cooling and then decreases with lowering temperature.
This is a common behavior of the magnetic part of $\rho(T)$ in
metallic magnets with periodic non-collinear spin structures, as
observed e.g. in holmium \cite{45}. With increasing magnetic field
the sharp kink at $T_N$$=$3.45 K is shifted down to 2 K for $H$=32.5
kOe (fig.\hyperref[FigX2]{2a}). This matches the shift of Neel temperature observed
in heat capacity data \cite{42} and the magnetic susceptibility
results of this compound \cite{43}. The magnetoresistance ratio
$\Delta\rho/\rho$$=$$[\rho(H)$$-$$\rho(H$$=$$0)]/\rho(H$$=$$0)$ for
Ho$_{0.5}$Lu$_{0.5}$B$_{12}$ is $-19\%$ (negative MR) at 3.45 K and
$H$=50 kOe. Moreover, an additional anomaly in resistivity can be
observed just below Neel temperature in low magnetic fields $H$$<$8
kOe (fig.\hyperref[FigX2]{2b}) which may be attributed to field dependent
spin-orientation magnetic phase transition, similar to that
detected inside the AF-phase of HoB$_{12}$ (Ref. \cite{46}).

\begin{figure*}
\begin{center}
\includegraphics[width = 10cm]{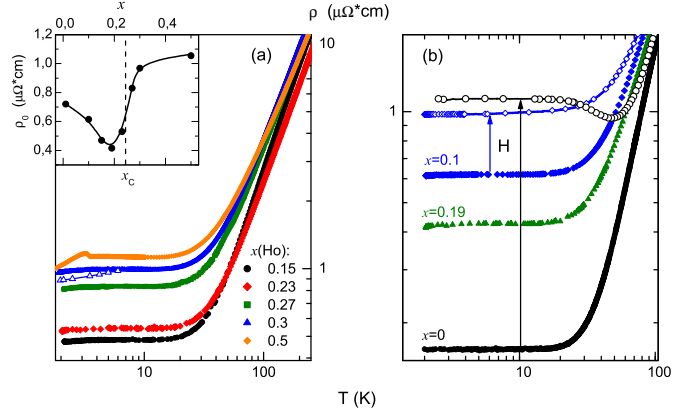}
   \caption{(a-b) The temperature dependences of electrical resistivity $\rho(T)$ of solid solutions Ho$_x$Lu$_{1-x}$B$_{12}$ with $x$=0, 0.1, 0.15, 0.19, 0.23, 0.3, and 0.5. Additionally presented are data for magnetic fields $H$=30 kOe ($x$=0.3, labeled by {\tiny $\triangle$} symbols on panel (a)) and $H$=80 kOe ($x$=0, 0.1, labeled  by $\circ$, $\diamond$  symbols on panel (b)) are presented. The inset in panel (a) shows the residual resistivity $\rho_0$ versus the holmium concentration $x$ (see the text). }\label{FigX3}
\end{center}
\end{figure*}

Figure \hyperref[FigX3]{3} shows the temperature dependences of resistivity both for
diluted ($x$=0.1, 0.15 and 0.19) and concentrated ($x$=0.23,
0.27, 0.3 and 0.5) magnetic Ho$_x$Lu$_{1-x}$B$_{12}$ solid
solutions in the range 1.9-300 K. For comparison, the $\rho$(T)
curve of the non-magnetic counterpart LuB$_{12}$ (the
4\textit{f}$^{14}$ configuration of Lu ion corresponds to the case
of a completely filled 4\textit{f}-shell of the rare earth ion) is
also presented in this figure. As can be seen from Fig.\hyperref[FigX3]{3}, all
RB$_{12}$ compounds under investigation are good metals, and in
absence of external magnetic field their ratio $\rho(300K)/
\rho(10K)$ exceeds 10 and reaches a maximum values of about 70 for
nonmagnetic LuB$_{12}$. In the Ho - concentration range
$x$=0.04$-$0.19 the residual resistivity $\rho_0$ decreases in the
range 0.4$-$0.7 $\mu\Omega\cdot$cm (see the insert in fig.\hyperref[FigX3]{3}), but
between $x$=0.23 and 0.3 $\rho_0$ exhibits a step-like increase with
rising Ho concentration, which is supposed to be a consequence of
the emergence of an infinite magnetic cluster (percolation) of
Ho-ions (see fig.\hyperref[FigX1]{1f}), although the $\rho(T)$ behavior does not
change considerably (fig.\hyperref[FigX3]{3}). At intermediate temperatures
resistivity can be described by a power law dependence $\rho(T)$$
\sim$$T^{\alpha}$ with exponent $\alpha$ varying in the range
between 1.3 and 1.7, depending on holmium content. Fig.\hyperref[FigX3]{3} displays
also the $\rho$(T) dependencies of both LuB$_{12}$ and solid
solutions Ho$_x$Lu$_{1-x}$B$_{12}$ with $x$=0.1 in magnetic field of
$H$=80 kOe, and for $x$=0.3 at $H$=30 kOe. It is worth to note that when
external magnetic field is applied, the $\rho(T, H$$=$80 kOe)
dependence of LuB$_{12}$ demonstrates an unconventional increase of
resistivity with decreasing temperature below the cage-glass
transition at $T^{\star}$$\sim$60 K (fig.\hyperref[FigX3]{3}). At the same time, all
studied Ho$_x$Lu$_{1-x}$B$_{12}$ dodecaborides demonstrate a
positive magnetoresistance in strong magnetic field. Taking into
account that both negative (fig.\hyperref[FigX2]{2} $-$ $T$$<$9K, fig.\hyperref[FigX3]{3a} $-$ curve $H$=30
kOe for $x$=0.3) as well as positive (fig.\hyperref[FigX3]{3b} $-$ in $H$=80 kOe) MR
effects are observed on high quality single crystals of
Ho$_x$Lu$_{1-x}$B$_{12}$, it appears to be important to measure in
detail the magnetic field dependencies $\rho(H,T_0)$ in a wide
range of temperatures, to separate and classify the
magnetoresistance contributions in these magnetic solid solutions
with various holmium content.

\subsection*{\emph{IIIb. Magnetoresistance of Ho$_x$Lu$_{1-x}$B$_{12}$.}}

\begin{figure*}
\begin{center}
\includegraphics[width = 10cm]{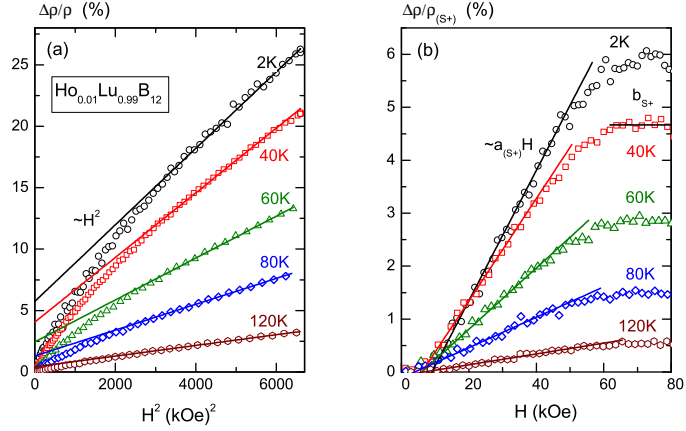}
   \caption{The field dependences of (a) magnetoresistance in coordinates $\Delta\rho(H)/\rho$$=$$f(H^2, T_0)$ and (b) $\Delta\rho(H)/\rho_{(S+)}$ the positive MR component of Ho$_{0.01}$Lu$_{0.99}$B$_{12}$ compound. The solid lines represent the (a) quadratic asymptotic $\sim$$H^2$, (b) linear $a_{(S+)}H$ and saturated $b_{(S+)}$ contributions to MR (see the text).}\label{FigX4}
   \end{center}
\end{figure*}

\begin{figure}[t]
\begin{center}
\subfigure{\includegraphics[scale=0.67]{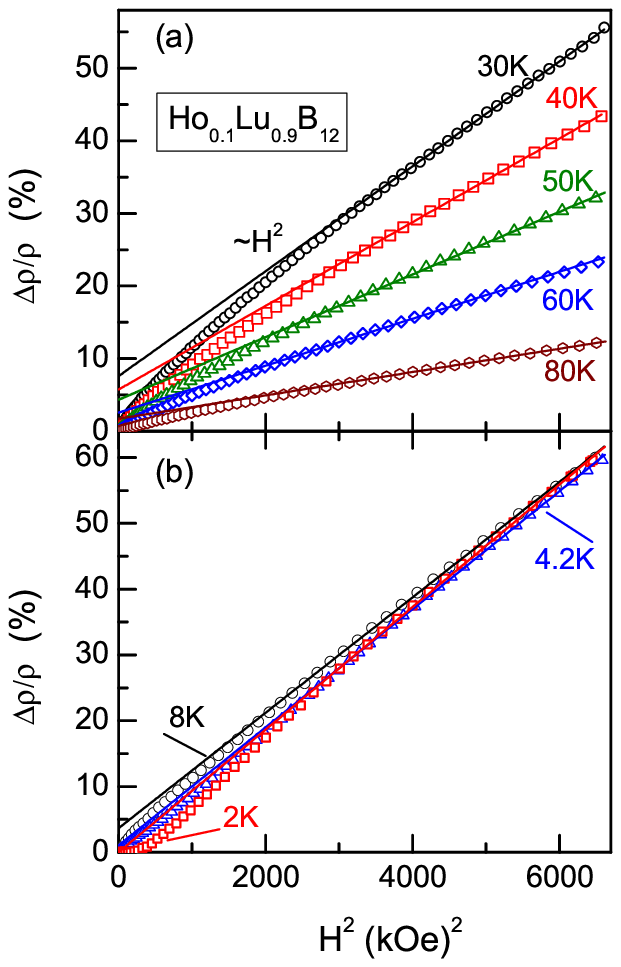}}
\subfigure{\includegraphics[scale=0.67]{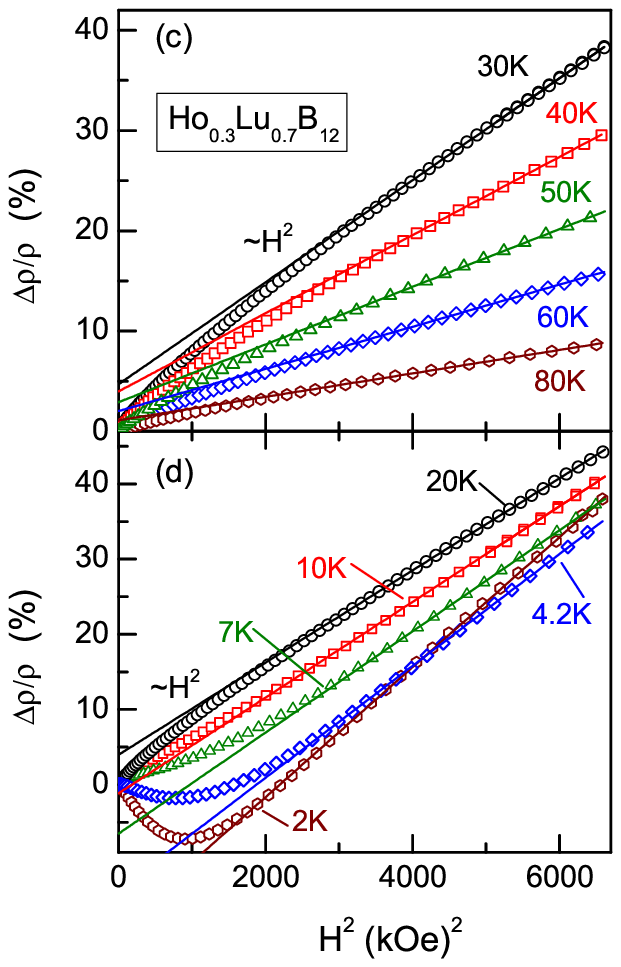}}
\caption{The field dependences of magnetoresistance  in coordinates $\Delta\rho(H)/\rho$$=$$f(H^2, T_0)$ of (a-b) Ho$_{0.1}$Lu$_{0.9}$B$_{12}$ and of (c-d) Ho$_{0.3}$Lu$_{0.7}$B$_{12}$ solid solutions. The solid lines in (a-d) represent the quadratic asymptotic $\sim$$H^2$.}\label{FigX5}
\end{center}
\end{figure}

\begin{figure}[t]
\begin{center}
\subfigure{\includegraphics[scale=0.7]{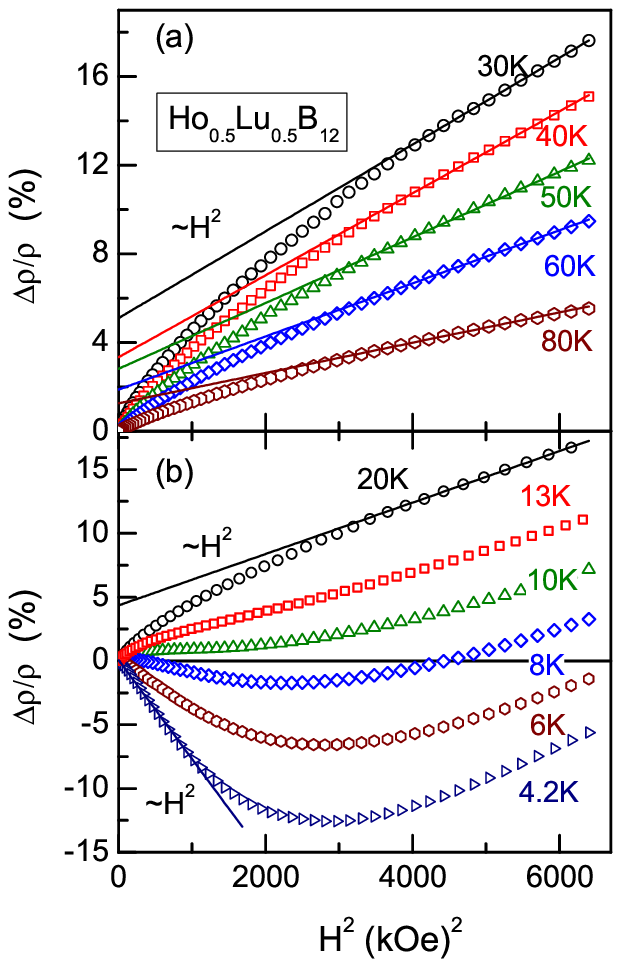}}
\subfigure{\includegraphics[scale=0.67]{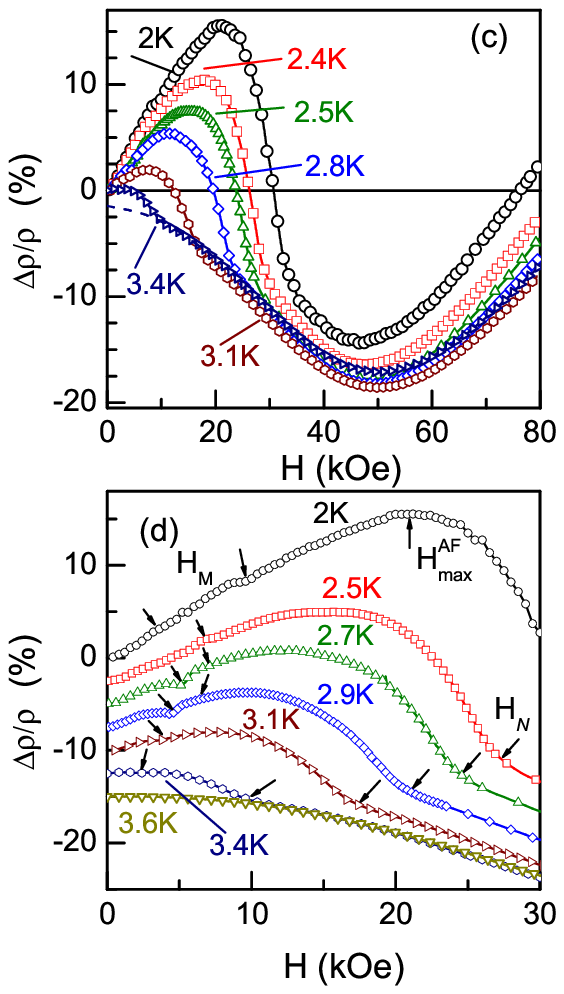}}
\parbox[t]{0.47\textwidth}{}
\,\parbox[t]{0.47\textwidth}{} \vspace{0.5cm} \caption{The field dependences of magnetoresistance in coordinates (a-b) $\Delta\rho(H)/\rho$$=$$f(H^2, T_0)$ and (c, d) $\Delta\rho(H)/\rho$$=$$f(H, T_0)$ of Ho$_{0.5}$Lu$_{0.5}$B$_{12}$ compound. The solid lines in (a-b) represent the quadratic asymptotic $\sim$$H^2$. The $\Delta\rho(H)/\rho$ curves in (d) are shifted by a constant value $\Delta\rho/\rho$=2.5 $\%$ for convenience. Arrows in (d) indicate the magnetic phase transitions (see the text).}\label{FigX6}
\end{center}
\end{figure}

\begin{figure*}[h]
\begin{center}
\includegraphics[width = 6cm]{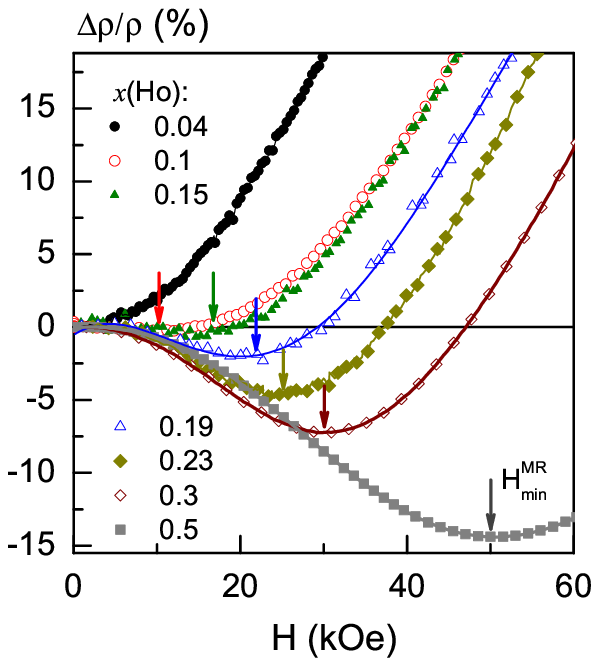}
   \caption{The field dependences of magnetoresistance of solid solutions Ho$_x$Lu$_{1-x}$B$_{12}$ with $x$=0.04, 0.1, 0.15, 0.19, 0.23, 0.3 ($T_0$$=$2 K) and 0.5 ($T_0$$=$3.6 K). The arrows indicate the positions of pronounced negative minima on MR data.}\label{FigX7}
 \end{center}
\end{figure*}

Results of MR investigations of Ho$_x$Lu$_{1-x}$B$_{12}$ solid
solutions with a Ho content $x$=0.01, 0.1, 0.3 and 0.5 are shown in
figs.\hyperref[FigX4]{4a}, \hyperref[FigX5]{5a-b}, \hyperref[FigX5]{5c-d}
 and \hyperref[FigX6]{6a-d}, correspondingly. As can be seen
from fig.\hyperref[FigX4]{4a}, in case of the diluted magnetic solid solution
Ho$_{0.01}$Lu$_{0.99}$B$_{12}$ the magnetoresistance is positive
anywhere and demonstrates a strong increase without a tendency to
saturation in high magnetic fields up to 80 kOe. With increase of
Ho content in the range $x$=0.1$-$0.5 the aforementioned positive MR
effect dominates in the range of intermediate temperatures $T$$>$20K
(see figs.\hyperref[FigX5]{5a}, \hyperref[FigX5]{5c}, \hyperref[FigX6]{6a}
  for $x$=0.1, 0.3 and 0.5, correspondingly),
but below 20 K an emergence of a negative contribution may be
observed on $\Delta\rho/\rho(H)$ curves, even for $x$=0.1 (fig.\hyperref[FigX5]{5b}).
At higher Ho concentration, in the range of $x$=0.19$-$0.5, a
pronounced negative minimum appears on MR \textit{vs} magnetic field
dependences at liquid helium temperatures (see fig.\hyperref[FigX7]{7} and also
figs.\hyperref[FigX5]{5d} and \hyperref[FigX6]{6b}) and its amplitude and location are strongly
dependent on $x$. Indeed, the amplitude of negative MR increases
essentially in the range $x$=0.15$-$0.5 (fig.\hyperref[FigX7]{7}) where the position of
MR minimum changes from $H_{min}^{MR}(x$$=$$0.15)$$\sim$16 kOe to
$H_{min}^{MR}(x$$=$$0.5)$$\sim$51 kOe (see also figs.\hyperref[FigX5]{5d} and \hyperref[FigX6]{6b-c}). For
comparison, a negative transverse MR with a large amplitude
($\sim$20$-$30$\%$) was observed previously \cite{41} in the
paramagnetic phase of HoB$_{12}$ at temperatures 7.5-15 K.
Moreover, the minimum value of negative magnetoresistance in
HoB$_{12}$ is expected to be observed in magnetic fields above 80
kOe \cite{41}.

The AF-P transition in Ho$_{0.5}$Lu$_{0.5}$B$_{12}$ which is,
depending on field, observed in the temperature range 1.9-3.5 K,
is accompanied by the appearance of an additional positive
antiferromagnetic MR contribution (see fig.\hyperref[FigX2]{2} and fig.\hyperref[FigX6]{6c,d}) which
becomes fully suppressed by external magnetic field when H reaches
the critical values $H_N(T_N)$ on the AF-P phase boundary
(fig.\hyperref[FigX6]{6d}). This additional positive MR component can be attributed
to charge carrier scattering on the magnetic structure in the AF
phase of Ho$_{0.5}$Lu$_{0.5}$B$_{12}$. A similar positive
magnetoresistance was observed previously in the AF phases
of HoB$_{12}$, ErB$_{12}$ and TmB$_{12}$ dodecaborides \cite{41} and
in antiferromagnetic solid solutions Tm$_{1-x}$Yb$_x$B$_{12}$ with
$x$$\leq$0.1 (Ref.\cite{47}). The mechanism which is responsible for
this effect will be discussed below.

\subsection*{\emph{IIIc. Hall effect and magnetic susceptibility.}}

\begin{figure*}[h]
\begin{center}
\includegraphics[width = 6cm]{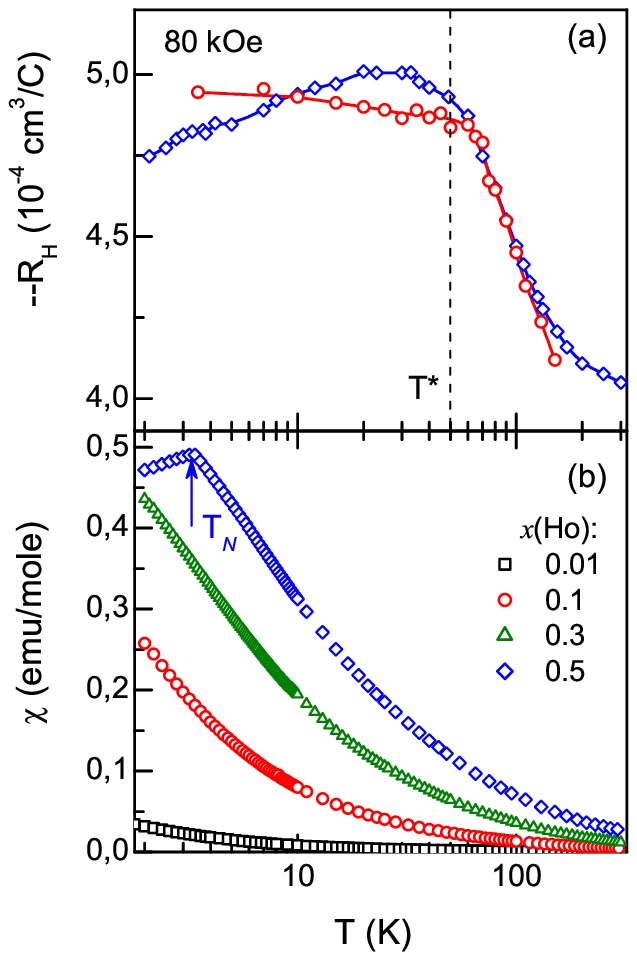}
   \caption{The temperature dependences of (a) negative Hall coefficient $-R_H(T)$ and (b) magnetic susceptibility $\chi(T)$ of solid solutions Ho$_x$Lu$_{1-x}$B$_{12}$ with $x$=0.01, 0.1, 0.3 and 0.5 (symbols {\tiny $\square$}, $\circ$, {\tiny $\triangle$}, $\diamond$, respectively).}\label{FigX8}
   \end{center}
\end{figure*}

Fig.\hyperref[FigX8]{8a} displays results of Hall effect measurements that have been
carried out simultaneously with magnetoresistance on several
crystals of Ho$_x$Lu$_{1-x}$B$_{12}$ at magnetic field $H$=80 kOe. A
pronounced increase (by 15-25$\%$) of the amplitude of negative
Hall coefficient is observed with temperature lowering in the
range above the cage-glass transition $T^{\star}$$\sim$60 K for all studied
Ho contents $x$=0.1 and 0.5 (fig.\hyperref[FigX8]{8a}). Then, a moderate
decrease of the absolute values of $R_H(T)$ can be observed with
temperature lowering below $T^{\star}$ for
Ho$_x$Lu$_{1-x}$B$_{12}$ compound with $x$=0.1
(fig.\hyperref[FigX8]{8a}). But on the contrary, for the most concentrated magnetic
dodecaboride with $x$=0.5 a moderate elevation of the negative
$R_H(T)$ is observed at $T$$<$$T^{\star}$ with a smooth maximum at
$T_{max}$$\sim$25 K. A similar strong field maximum of negative Hall
coefficient was found at 20 K in \cite{48} for HoB$_{12}$.

The magnetic susceptibility $\chi(T)$ dependencies received in the
present study in small magnetic fields $H$=0.1 kOe (for $x$=0.1, 0.3
and 0.5) and in $H$=5 kOe (for $x$=0.01) at temperatures in the range
2-300 K are shown in fig.\hyperref[FigX8]{8b}. The $\chi(T)$ dependences
demonstrate a paramagnetic Curie-Weiss type behavior, and for
Ho$_x$Lu$_{1-x}$B$_{12}$ solid solution with $x$=0.5 the AF-P phase
transition is observed at $T_N$$\approx$3.45 K (fig.\hyperref[FigX8]{8b}).

In the further analysis and discussion of these results it will be
shown that it is possible to separate and classify the
aforementioned positive and negative contributions to
magnetoresistance. Moreover, the developed approach will
demonstrate that the MR components may be estimated quantitatively
and that parameters extracted from the data will allow to describe
both the temperature dependence of drift mobility and the
characteristics of nanosize magnetic clusters (domains with
AF-type short range order) which are composed from interconnected
Ho$^{3+}$-ions embedded in the rigid covalent cage of boron atoms.

\section*{IV. DISCUSSION}

To separate and characterize analytically the large negative
magnetoresistance effect observed in vicinity of AF-P transition
in Ho$_{0.5}$Lu$_{0.5}$B$_{12}$ (fig.\hyperref[FigX2]{2}) it is first necessary to
analyze several positive MR components which are dominant
(\textit{i}) at intermediate temperatures 20-100 K and
(\textit{ii}) in the antiferromagnetic state of
Ho$_x$Lu$_{1-x}$B$_{12}$ compounds. The most effective approach
here is based on investigations of the positive MR in the
paramagnetic state of diluted Ho$_x$Lu$_{1-x}$B$_{12}$ solid
solutions (with a small concentration of magnetic impurities
$x$=0.01 and 0.1). From the side of percolation threshold (see inset
in fig.\hyperref[FigX3]{3a}), for $x$=0.3 and 0.5, a large negative MR component
appears at low temperatures ($T_N$$\leq$$T$$\leq$10 K) versus the
positive MR background, and these terms should be analyzed in
combination with each other. In the approximation of several
additive processes in the scattering of charge carriers in
Ho$_x$Lu$_{1-x}$B$_{12}$, we will gradually develop a
phenomenological approach to separate of the MR contributions, and
then propose the interpretation of the large negative
magnetoresistance. We will demonstrate that large negative MR
effect may be explained by scattering of charge carriers on the
small size magnetic clusters of Ho-ions, and the Yosida-type
relationship $-\Delta\rho/\rho$$\sim$$L^2(H/T)$ (\textit{L} - Langevin
function) will provide us with a good quality approximation of the
negative magnetoresistance behavior. Simultaneously, the analysis
of MR in the AF phase will allow us to conclude in favor of a
combination of \textit{5d} and 4\textit{f}- components which
interplay with each other in the formation of the complicated
magnetic structure of Ho$_{0.5}$Lu$_{0.5}$B$_{12}$.

\subsection*{\emph{IVa. Positive magnetoresistance in the paramagnetic phase of Ho$_x$Lu$_{1-x}$B$_{12}$.}}

\begin{figure*}
\begin{center}
\includegraphics[width = 6cm]{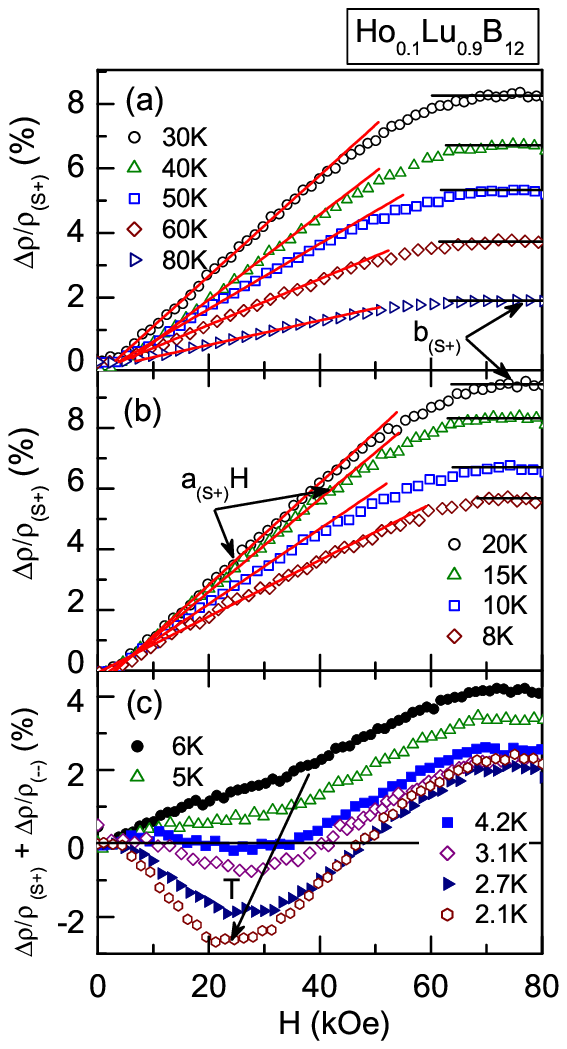}
   \caption{The field dependences of (a-b) $\Delta\rho/\rho_{(S+)}$ and (c) ($\Delta\rho/\rho_{(S+)}+\Delta\rho/\rho_{(-)}$) contributions to MR of Ho$_{0.1}$Lu$_{0.9}$B$_{12}$ compound. The solid lines correspond to the positive linear $a_{(S+)}H$ (red) and  saturated $b_{(S+)}$ (black) components of MR term $\Delta\rho/\rho_{(S+)}$, respectively.}\label{FigX9}
 \end{center}
\end{figure*}

\begin{figure*}
\begin{center}
\includegraphics[width = 10cm]{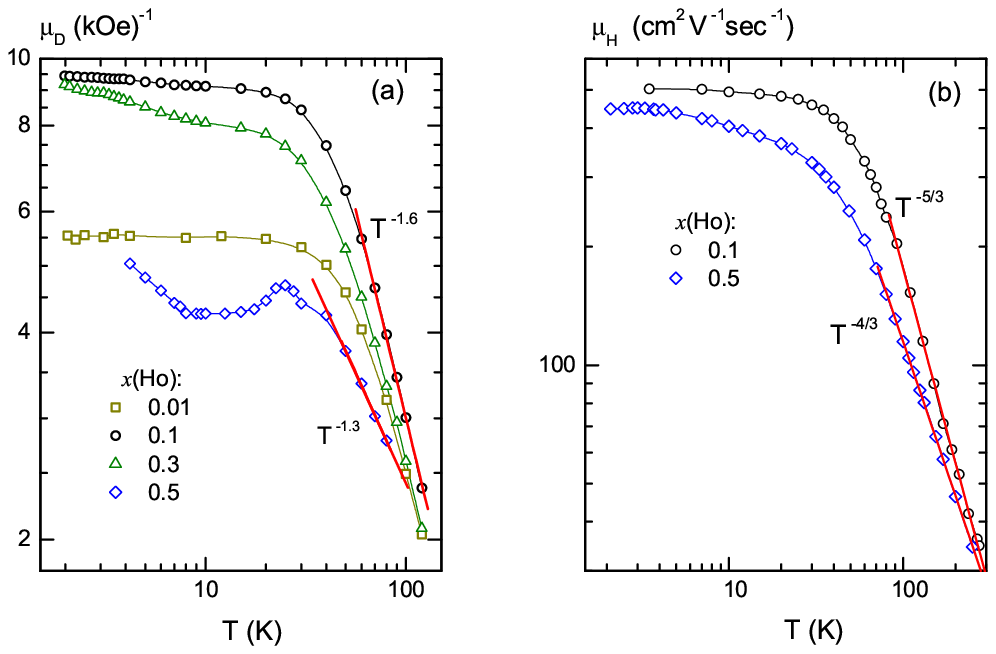}
   \caption{The temperature dependences of (a) the reduced drift mobility of charge carriers $\mu_D(T)$ and (b) Hall mobility $\mu_H(T)$$=$$R_H(T)/\rho(T)$ of solid solutions Ho$_x$Lu$_{1-x}$B$_{12}$ with $x$=0.01, 0.1, 0.3 and 0.5 (symbols {\tiny $\square$}, $\circ$, {\tiny $\triangle$}, $\diamond$, respectively). The solid lines on both panels represent the data approximation by the power law $\mu$$\sim$$T^{-\alpha}$ (see the text).}\label{FigX10}
 \end{center}
\end{figure*}

Both the MR data presented in coordinates $\Delta\rho/\rho$$=$$
f(H^2, T_0)$ (see solid lines in figs.\hyperref[FigX4]{4a}, \hyperref[FigX5]{5a}, \hyperref[FigX5]{c} and \hyperref[FigX6]{6a}, \hyperref[FigX6]{b}) and
their numerical derivatives ($d(\Delta\rho/\rho)/dH$$=$$f(H, T_0)$
(see Fig.\hyperref[FigXS1]{S1} in the \hyperref[SI]{Supplementary Information}) reveal that the
positive magnetoresistance observed at intermediate temperatures
in high magnetic fields 50-80 kOe follows the quadratic field
dependence $\Delta\rho/\rho_{(m+)}$$=$$\mu_D^2H^2$, where from
conventional approach the parameter $\mu_D$ may be considered as
the reduced drift mobility of charge carriers. Additionally to
this dominant quadratic term $\Delta\rho/\rho_{(m+)}$, there is
another positive component $\Delta\rho/\rho_{(s+)}$ detected in a
moderate magnetic fields. This second positive MR contribution may
be singled out at intermediate temperatures by subtracting the
strong field term $\Delta \rho/\rho_{(m+)}$$=$$\mu_D^2H^2$ from the
observed experimental $\Delta\rho/\rho(H)$ dependence. The final
$\Delta\rho/\rho_{(s+)}$  is shown, for example, in figs.\hyperref[FigX4]{4b} and
\hyperref[FigX9]{9a}, \hyperref[FigX9]{b} for holmium content $x$=0.01 and 0.1, respectively. It can be
seen from these figures that the second term is negligible in
small magnetic fields $H$$\leq$5 kOe, but it demonstrates an
approximately linear field dependence $\Delta\rho/\rho_{(s+)}$$\sim$$a_{(s+)}H$
 in the range of 10-40 kOe. Then it saturates at high
magnetic fields (figs.\hyperref[FigX4]{4b}, \hyperref[FigX9]{9a}) and its amplitude $b_{(s+)}$ does
not exceed 9$\%$ at $T$$>$7K for all Ho$_x$Lu$_{1-x}$B$_{12}$
crystals studied. Moreover, in case of the diluted magnetic system
Ho$_{0.01}$Lu$_{0.99}$B$_{12}$ the magnetoresistance exhibits only
these two positive contributions both at intermediate and liquid
helium temperatures (fig.\hyperref[FigX4]{4b}). In the absence of the negative MR
the analysis of two positive terms $\Delta\rho/\rho_{(m+)}$ and
$\Delta\rho/\rho_{(s+)}$ allows to deduce the temperature
dependences of the above-mentioned coefficients $\mu_D$,
$a_{(s+)}$ and $b_{(s+)}$ for $x$=0.01, and also for $x$$\geq$0.1 in
the range $T$$>$10 K. The resulting $\mu_D$(T) dependence acquired
directly from data of figs.\hyperref[FigX4]{4}-\hyperref[FigX6]{6} and fig.\hyperref[FigX9]{9} is shown on fig.\hyperref[FigX10]{10a}.
Additionally, fig.\hyperref[FigX10]{10b} shows for comparison the high field Hall
mobility $\mu_H(T)$$=$$R_H(T)/\rho(T)$ for Ho$_x$Lu$_{1-x}$B$_{12}$
with $x$=0.1 and 0.5. It can be seen that the behavior of reduced
drift $\mu_D$ and Hall $\mu_H$ mobility at $H$=80 kOe is similar to
each other, and that exponents $\alpha_D$ and $\alpha_H$ in
dependences $\mu_D(T)$$\sim$$T^{-\alpha_D}$ and $\mu_H(T)$$\sim$$T^{
-\alpha_H}$ obtained at intermediate temperatures $T$$\geq
$$T^{\star}$$\sim$60 K are about equal in studied compounds
($\alpha_D(x$$=$$0.1)$$\approx$$\alpha_H(x$$=$$0.1)$$\approx$1.6-1.7 and
$\alpha_D(x$$=$$0.5)$$\approx$$\alpha_H(x$$=$$0.5)$$\approx$1.3, see fig.\hyperref[FigX10]{10}).
The deduced temperature dependences of coefficient $a_{(s+)}$ and
the saturation value $b_{(s+)}$ of the $\Delta\rho/\rho_{(s+)}$
contribution of all compounds (see fig.\hyperref[FigXS2]{S2} in the \hyperref[SI]{SI}) are also
similar and connect the slope of the linear increase of MR with
the amplitude of the second positive MR component of
Ho$_x$Lu$_{1-x}$B$_{12}$ solid solutions.

\subsection*{\emph{IVb. Negative magnetoresistance in the paramagnetic phase of Ho$_x$Lu$_{1-x}$B$_{12}$.}}

\begin{figure*}
\begin{center}
\includegraphics[width = 5cm]{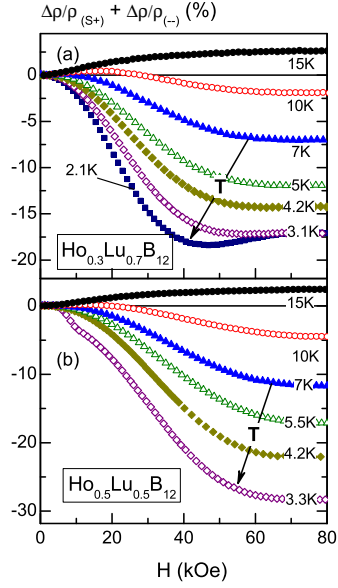}
   \caption{The field dependences of $\Delta\rho/\rho_{(S+)}+\Delta\rho/\rho_{(-)}$ contribution to MR (see text) for (a) Ho$_{0.3}$Lu$_{0.7}$B$_{12}$ and (b) Ho$_{0.5}$Lu$_{0.5}$B$_{12}$ compounds.}\label{FigX11}
   \end{center}
\end{figure*}

Apart from looking for the physical meaning of the second
$\Delta\rho/\rho_{(s+)}$ component, it should be stressed that the
simple phenomenological procedure applied above for evaluating the
positive MR may be developed successfully also to separate the
negative magnetoresistance observed for Ho$_x$Lu$_{1-x}$B$_{12}$
with $x$$\geq$0.1 at low temperatures. Indeed, on the contrary to
the dilute compound with $x$=0.01 where the only two positive MR
components with coefficients $\mu_D$ and $a_{(s+)}$, $b_{(s+)}$
have been observed in a wide temperature range 1.9$-$120 K (see
fig.\hyperref[FigX10]{10a} and fig.\hyperref[FigXS2]{S2}), for magnetic dodecaborides
Ho$_x$Lu$_{1-x}$B$_{12}$ with $x$$\geq$0.1 the emergence of
additional negative magnetoresistance is evident at low
temperatures $T$$<$10 K, and its amplitude increases dramatically
with temperature lowering (see figs.\hyperref[FigX9]{9c} and \hyperref[FigX11]{11a}, \hyperref[FigX11]{b}). Moreover, in
Ho$_x$Lu$_{1-x}$B$_{12}$ solid solutions the increase of Ho
content is accompanied by a strong increase of the negative MR
contribution, e.g. between $\Delta\rho/\rho_{(-)}(x$$=$$0.1)$$\sim$3 $\%$
(fig.\hyperref[FigX9]{9c}) and $\Delta\rho/\rho_{(-)}(x$$=$$0.5)$$\sim$30 $\%$ (fig.\hyperref[FigX11]{11b}). A
comparative analysis of data presented in figs.\hyperref[FigX9]{9c} and \hyperref[FigX11]{11} allows
to conclude that (\textit{i}) at low magnetic field the negative
MR follows a quadratic dependence $-\Delta\rho/\rho_{(-)}$$\sim$$H^2$
and (\textit{ii}) that the $\Delta\rho/\rho_{(-)}$ term
demonstrates also a tendency to saturation in high magnetic
fields. It should be mentioned that such kind of behavior of
$\Delta\rho/\rho_{(-)}$ is well-known both for manganites
\cite{49}, non-magnetic and AF heavy fermion compounds like
CeCu$_6$, CeAl$_3$ \cite{50} and CeAl$_2$, CeB$_6$ \cite{34, 51},
 AF metal GdSi \cite{32}, etc. Recently this effect has
been observed also in dodecaborides HoB$_{12}$, ErB$_{12}$ and
TmB$_{12}$ \cite{41} and hexaborides PrB$_6$, NdB$_6$ and GdB$_6$
\cite{52, 53}, and it was analyzed successfully within the
framework of Yosida approach \cite{54} based on \textit{s-d}
exchange model. This model describes the scattering of charge
carries on localized magnetic moments (LMM) by the relationship
between negative MR and local magnetization \textbf{M}$_{loc}$

\begin{equation}\label{Eq.1}
    -\Delta\rho/\rho_{(-)} \sim M_{loc}^2 .
\end{equation}
In small magnetic fields, where the linear dependence
\textbf{M}$_{loc}$$\sim$$\chi_{loc}\textbf{H}$ is valid, the relationship
(\hyperref[Eq.1]{1}) allows to explain a simple quadratic field dependence
of negative magnetoresistance. Moreover, it was shown in
\cite{50}-\cite{53} that the behavior of local magnetic
susceptibility $\chi_{loc}(T)$ may be detected directly from the
study of the $\Delta\rho/\rho_{(-)}$ term. The emergence of strong
negative MR in heavy fermion compounds was attributed \cite{34, 50, 51} to a formation of spin-polaron resonance in
the electron density of states (DOS) at $E_F$, which appears in
systems with strong local 4\textit{f-5d} spin fluctuations.
Simultaneous polarization of R$^{3+}$ magnetic moments of rare
earth ions and of the spins of conduction electrons in external
magnetic field destroys the DOS resonance and prevents the on-site
spin-flip scattering.

\begin{figure*}
\begin{center}
\includegraphics[width = 5cm]{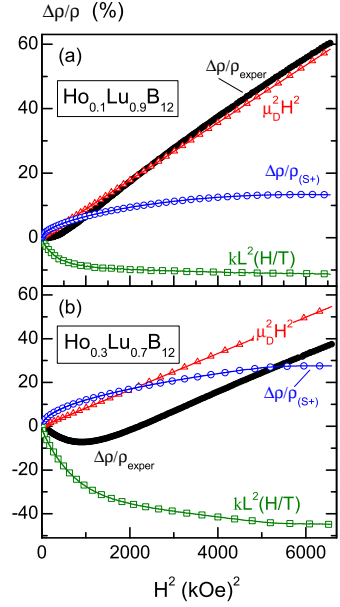}
   \caption{Two examples of MR analysis with contribution separation for (a) Ho$_{0.1}$Lu$_{0.9}$B$_{12}$ and (b) Ho$_{0.3}$Lu$_{0.7}$B$_{12}$ compounds at $T_0$$=$2 K. Symbols: $\bullet$ correspond to  experimental data ($\Delta\rho/\rho_{exper}$), {\tiny$\triangle$}, $\circ$ $-$ to positive contributions: $\mu_{D}^2H^2$ and $\Delta\rho/\rho_{(S+)}$ and {\tiny $\square$} $-$ to the saturated magnetic component $kL^2(H/T)$, respectively.}\label{FigX12}
 \end{center}
\end{figure*}

Taking into account the formation of nanosize magnetic clusters of
holmium ions in the dodecaboride matrix (see fig.\hyperref[FigX1]{1e}, \hyperref[FigX1]{f}), it is
natural to expect that the Langevin function
$L(\alpha)$$=$$cth(\alpha)$$-$$1/\alpha$ (where $\alpha$$=$$\mu_{eff}H/k_BT$
, $k_B$ is the Boltzmann constant and $\mu_{eff}$ the effective
magnetic moment of the magnetic nanodomains) should provide an
appropriate approximation of the local magnetization behavior. As
a result, the sum of three terms $\Delta\rho/\rho_{(m+)}$$=$$
\mu_D^2H^2$, $\Delta\rho/\rho_{(s+)}$$=$$(a_{(s+)}H; b_{(s+)})$ and
$\Delta\rho/\rho_{(-)}$$=$$kL^2(\alpha)$ was taken to fit the
magnetoresistance at $T$$<$10 K in the paramagnetic phase of
Ho$_x$Lu$_{1-x}$B$_{12}$ solid solutions. In the first step of the
procedure the coefficient $\mu_D$ was received directly from the
approximation of high field experimental data
$\Delta\rho/\rho(H,T_0)$ (see e.g. figs.\hyperref[FigX5]{5b}, \hyperref[FigX5]{d} and \hyperref[FigX6]{6a}). Parameters
$a_{(s+)}(T)$, $b_{(s+)}(T)$ and $\mu_{eff}(T)$ have been deduced
from the analysis of residual MR data analysis. For example, figures \hyperref[FigX12]{12a} and
\hyperref[FigX12]{12b} show the received contributions to magnetoresistance-
$\mu_D^2H^2$, $\Delta\rho/\rho_{(s+)}$ and $kL^2(H/T_0)$
simultaneously with the experimental curves of $\Delta\rho/\rho(H,
T_0)$ recorded at $T$$\sim$2 K for $x=$0.1 and $x=$0.3,
respectively.

\begin{figure*}
\begin{center}
\includegraphics[width = 10cm]{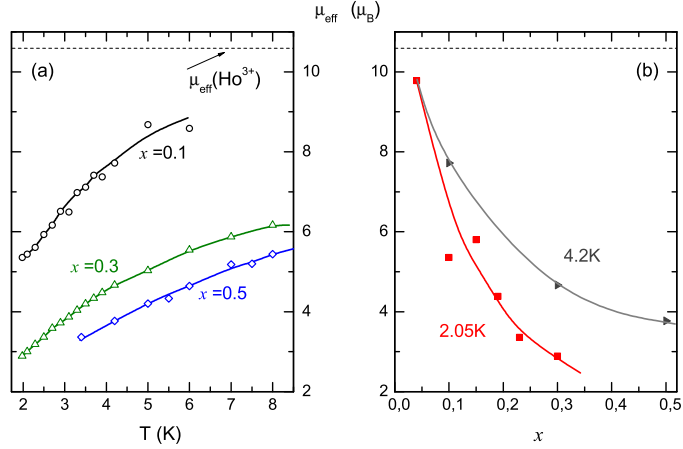}
   \caption{The temperature (a) and the concentration (b) dependences of effective magnetic moment $\mu_{eff}$ of Ho$_{x}$Lu$_{1-x}$B$_{12}$ with $x$$\leq$0.5 (see the text). The dashed lines on both panels display the value of effective magnetic moment of Ho$^{3+}$ ion.}\label{FigX13}
   \end{center}
\end{figure*}

\begin{figure*}
\begin{center}
\includegraphics[width = 10cm]{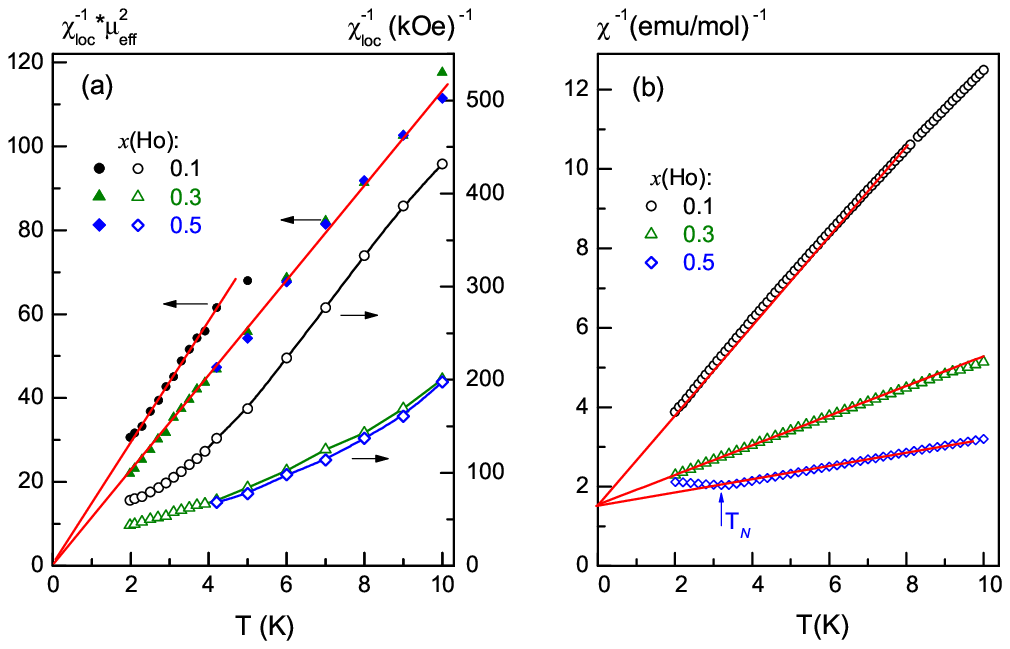}
   \caption{The temperature dependences of (a) the product $\chi_{loc}^{-1}$$\cdot$$\mu_{eff}^2$ (left axis) and the reciprocal local magnetic susceptibility $\chi_{loc}^{-1}(T)$ (right axis) and (b) reciprocal bulk magnetic susceptibility $\chi^{-1}(T)$ for Ho$_x$Lu$_{1-x}$B$_{12}$ solid solutions with $x$=0.1, 0.3 and 0.5. The solid straight lines show linear approximation.}\label{FigX14}
  \end{center}
\end{figure*}

Parameters $\mu_{eff}(T,x)$ and $\chi_{loc}^{-1}(T,x_0)$ obtained
in the framework of this approach are presented in figs.\hyperref[FigX13]{13} and
\hyperref[FigX14]{14a}, correspondingly. For comparison, panel \hyperref[FigX14]{b} of fig.\hyperref[FigX14]{14} displays
also reciprocal bulk magnetic susceptibility data
$\chi^{-1}(T,x_0)$ recorded for the same Ho$_x$Lu$_{1-x}$B$_{12}$
crystals. It is worth to note that the reduced local
susceptibility  $\chi_{loc}(T,x_0)$$=$$1/H\cdot
(d(-\Delta\rho/\rho)/dH)^{-1/2}$  was deduced directly from small
field ($H$$<$5 kOe) MR data $\Delta\rho/\rho_{exper} -
\Delta\rho/\rho_{(m+)}$, and the parameter $\mu_{eff}(T,x)$ was
independently determined by fitting of the sum
$\Delta\rho/\rho_{(s+)} + \Delta\rho/\rho_{(-)}$ (see figs.\hyperref[FigX9]{9c},
\hyperref[FigX11]{11a,b}) in two magnetic field ranges 10-40 kOe and 60-80 kOe.

For the system with nanosize magnetic clusters arranged from the
holmium ions (see fig.\hyperref[FigX1]{1e,f}) it is natural to expect a noticeable
reduction of $\mu_{eff}$ values in comparison with Ho$^{3+}$
magnetic moment $\mu(Ho^{3+})$$=$10.6 $\mu_B$.  Indeed, these small
size clusters with AF exchange interaction inside them may be
considered as nanoscale magnetic domains with AF short range
order. In these terms the small size clusters of Ho$^{3+}$ ions
with reduced LMM values should be treated as classical magnetic
moments with effective moment $\mu_{eff}$  whose magnetization is
described by the Langevin function $L(\alpha)$$=$$cth(\alpha)-
1/\alpha$ (where $\alpha$$=$$\mu_{eff}H/k_BT$). A decrease of
$\mu_{eff}$ is observed in this study both (\textit{i}) with the
temperature lowering (fig.\hyperref[FigX13]{13a}) and (\textit{ii}) with the
increase in Ho content (fig.\hyperref[FigX13]{13b}) in Ho$_x$Lu$_{1-x}$B$_{12}$
solid solutions. Thus, the AF short range order formation can be
considered as the most adequate interpretation of the effective
moment reduction. The Kondo mechanism of charge carriers
scattering cannot be responsible for the negative MR effect in
Ho-based dodecaborides, as Ho$^{3+}$ (4\textit{f}$^{10}$
configuration, $\Gamma_{51}$ triplet ground state) is not a Kondo-ion.
Moreover, the analysis of magnetoresistance of HoB$_{12}$,
ErB$_{12}$ and TmB$_{12}$ undertaken in \cite{41} allows
concluding that both spin-polaron and short range order effects
are responsible for the appearance of negative magnetoresistance
in these dodecaborides. So, the reduction of $\mu_{eff}(T)$ (fig.\hyperref[FigX13]{13a})
 and $\mu_{eff}(x)$ (fig.\hyperref[FigX13]{13b}) in paramagnetic state of these
cage-glass compounds with nanosize magnetic clusters may be
attributed directly to the extension of AF domains in the
RB$_{12}$ matrix. Moreover, following to the developed approach it
becomes possible to interpret also the violation of Curie-Weiss
law $\chi(T)$$\sim$$\mu_{eff}^2/(T-\theta_p)$ observed for all three
$\chi_{loc}(T)$ curves (fig.\hyperref[FigX14]{14a}, right axis). In this way, taking
into account the strong reduction of $\mu_{eff}$ with temperature
lowering (fig.\hyperref[FigX13]{13a}) one has to analyze the product
$\chi_{loc}^{-1}(T)$$\cdot$$\mu_{eff}^2(T)$ which is expected to follow the
Curie-Weiss relation. And indeed, a linear temperature dependence
of the product $\chi_{loc}^{-1}$$\cdot$$\mu_{eff}^2$ is observed for
Ho$_x$Lu$_{1-x}$B$_{12}$ crystals with negative MR - $x$=0.1, 0.3 and
0.5 (see fig.\hyperref[FigX14]{14a}, left axis). Finally, it should be stressed that
for concentrated holmium compounds with $x$=0.3 and $x$=0.5 we
have obtained almost equal values of  $\chi_{loc}^{-1}(T)$
(fig.\hyperref[FigX14]{14a}) and similar values of $\mu_{eff}(T)$ (fig.\hyperref[FigX13]{13a}). These
results allow us to conclude that these finite size
antiferromagnetic nanodomains are responsible both for the
spin-flip scattering as well as for the appearance of negative MR
in the magnetic dodecaborides.

\subsection*{\emph{IVc. Magnetic phase diagram and MR contributions in the AF state of Ho$_{0.5}$Lu$_{0.5}$B$_{12}$.}}

\begin{figure*}
 \begin{center}
\includegraphics[width = 8cm]{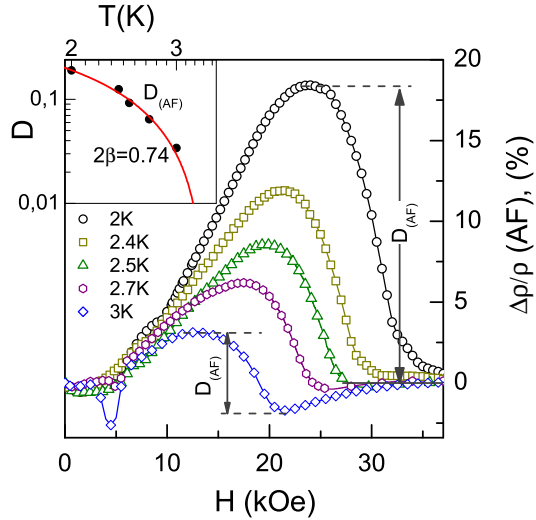}
   \caption{The field dependences of $\Delta\rho/\rho$(AF) term of magnetoresistance of Ho$_{0.5}$Lu$_{0.5}$B$_{12}$. $D_{AF}$ designate the amplitude of the MR component. The inset shows the result of critical exponent analysis $D_{AF}$$\sim$$(1-T/T_N)^{-2\beta}$ (see the text).}\label{FigX15}
    \end{center}
\end{figure*}

As it was mentioned above, the MR effect in the AF phase of
Ho$_x$Lu$_{1-x}$B$_{12}$ is quite different from that observed in
the paramagnetic state. Indeed, at temperatures $T$$<$$T_N$ new large
positive magnetoresistance component becomes dominant in moderate
magnetic fields, but, in the range $H_{max}^{AF}$$<$$H$$<$$H_N$ a strong
MR decrease can be seen (fig.\hyperref[FigX6]{6c,d}). To estimate quantitatively the
amplitude of the positive $\Delta\rho/\rho$(AF) term observed in
the AF phase we have used an extrapolation of the sum of
paramagnetic contributions observed in the P phase in magnetic
fields $H$$<$$H_N$ (see fig.\hyperref[FigX6]{6c}), then the $\Delta\rho/\rho$(AF) part
can be deduced by subtracting this paramagnetic background from
experimental MR data. The obtained AF contribution to MR in
Ho$_{0.5}$Lu$_{0.5}$B$_{12}$ is shown in fig.\hyperref[FigX15]{15}. Following to the
analysis of critical behavior of MR developed in \cite{41}, we
have calculated also the critical exponent for the amplitude
$D_{(AF)}$ of $\Delta\rho/\rho$(AF) (see fig.\hyperref[FigX15]{15}) in the framework
of relation $D_{(AF)}$$\sim$$(1- T/T_N)^{2\beta}$. The exponent
$\beta$$=$0.37$\pm$0.02 calculated for Ho$_{0.5}$Lu$_{0.5}$B$_{12}$
(see inset in fig.\hyperref[FigX15]{15}) agrees very well with values $\beta$$=$0.36
and 0.43 received previously in \cite{41} for HoB$_{12}$,
ErB$_{12}$ and TmB$_{12}$, respectively. Within the framework of
Yosida model \cite{54} (see eq.\ref{Eq.1}) a critical behavior of
magnetoresistance is expected in vicinity of $T_N$, and exponents
for MR ($\eta$) and for local magnetization ($\beta$) should be
connected by relation $\eta$$=$$2\beta$. The critical exponent
$\beta$$=$0.37 obtained here for $M_{loc}$$\sim$$(-\Delta\rho/\rho)^{1/2}$ is close to values $\beta$$=$
0.335$\pm$0.005 and $\beta$$=$0.385$\pm$0.01 previously observed in
magnetization studies of MnF$_2$ and RbMnF$_3$ antiferromagnets
\cite{55, 56}. In case of a three-dimensional Heisenberg model the
critical exponent of magnetization calculated by expanding into
series has a value of $\beta$$=$0.38$\pm$0.03, whereas the critical
exponent for the three-dimensional Ising model is $\beta$$=$0.312$\pm$0.005
 \cite{57}. Therefore, the value of $\beta$$=$0.37
obtained in this study for the Ho$_{0.5}$Lu$_{0.5}$B$_{12}$ magnet
is physically justified and, according to our opinion, it can
serve as an additional argument in favor of applicability of the
spin-polaron approach used here to describe the magnetoresistance
of the Ho$_x$Lu$_{1-x}$B$_{12}$ antiferromagnets. It is also worth to note
that the temperature dependence of both the amplitude $I_{AF}$ and
the width $\Delta_{AF}$ of magnetic Bragg maxima for the
antiferromagnetic phase of parent HoB$_{12}$ compound was recently
investigated in \cite{58} using neutron diffraction technique. It
was revealed that in vicinity of $T_N$ the parameters $I_{AF}$ and
$\Delta_{AF}$ are characterized by a critical behavior with
exponents $\beta$$\approx$$\gamma$$\approx$1/3, which are also in
accordance with the critical exponent  $\beta$$\approx$0.37
received here for Ho$_{0.5}$Lu$_{0.5}$B$_{12}$.

\begin{figure*}
 \begin{center}
\includegraphics[width = 15cm]{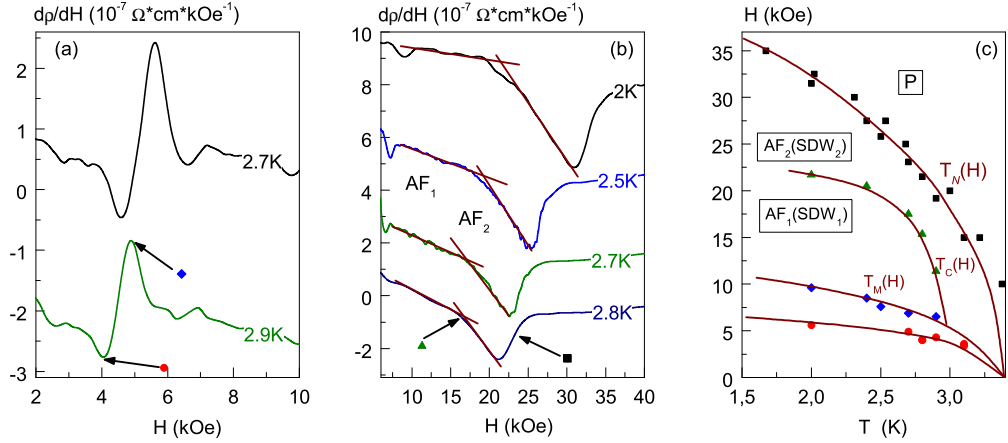}
   \caption{(a-b) Results of the numerical differentiation of resistivity $d\rho/dH$ in AF state of Ho$_{0.5}$Lu$_{0.5}$B$_{12}$ compound. The arrows indicate the magnetic phase transitions. (c) \textit{H-T} magnetic phase diagram of Ho$_{0.5}$Lu$_{0.5}$B$_{12}$, reconstructed from MR data. Abbreviations: P $-$ paramagnetic state, AF$_{1,2}$ (SDW$_{1, 2}$) $-$ antiferromagnetic spin-density wave phases.}\label{FigX16}
   \end{center}
\end{figure*}

To identify precisely the magnetic phase transitions both between
AF and P phases and inside the AF state, a numerical
differentiation analysis of resistivity curves has been carried
out, and features of derivatives $d\rho/dH$ (see figs.\hyperref[FigX16]{16a,b}) were
used to construct the \textit{H-T} magnetic phase diagram of
Ho$_{0.5}$Lu$_{0.5}$B$_{12}$ (see fig.\hyperref[FigX16]{16c}). The received
\textit{H-T} diagram presented in Fig.\hyperref[FigX16]{16c} for
Ho$_{0.5}$Lu$_{0.5}$B$_{12}$ is similar to that observed in
\cite{41, 48} for parent antiferromagnet HoB$_{12}$ with a higher
Neel temperature $T_N$$\approx$7.4 K. The high accuracy of
resistivity measurements allowed us also to analyze and classify
the MR components observed in the AF$_1$ and AF$_2$ phases of the solid
solution with $x$=0.5. Indeed, the analysis of linear fragments of resistivity
derivatives (fig.\hyperref[FigX16]{16b}) allows to describe the magnetorestistance of
Ho$_{0.5}$Lu$_{0.5}$B$_{12}$ through  the relationship

\begin{figure*}
 \begin{center}
\includegraphics[width = 5cm]{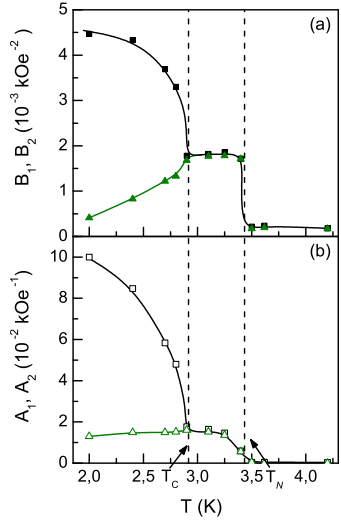}
   \caption{The temperature dependences of the amplitudes of (a) negative quadratic term ($B_{1, 2}$) and (b) linear positive ($A_{1, 2}$) component in magnetoresistance of Ho$_{0.5}$Lu$_{0.5}$B$_{12}$ compound (see the Eq.\ref{Eq.2}).}\label{FigX17}
   \end{center}
\end{figure*}

\begin{equation}\label{Eq.2}
   \Delta\rho/\rho(H,T_0) = -B_{1,2}(T_0)H^2 + A_{1,2}(T_0)H + C.
\end{equation}
Eq.\ref{Eq.2} provides us with a good quality approximation of the MR
results both below (phase AF$_1$, $B_1$ and $A_1$ coefficients in Eq.\ref{Eq.2})
 and above (phase AF$_2$, $B_2$ and $A_2$ coefficients
in Eq.\ref{Eq.2}) the $T_C(H)$ phase boundary (linear fits for
$d\rho/dH$ are shown in fig.\hyperref[FigX16]{16b}). The calculated temperature
dependences of coefficients $B_{1, 2}(T_0)$ and $A_{1, 2}(T_0)$ are
presented in figs.\hyperref[FigX17]{17a,b}, respectively. As can be seen from
figs.\hyperref[FigX16]{16b} and \hyperref[FigX17]{17b}, simultaneously with the negative quadratic term
$\Delta\rho/\rho_{(-)}$$=$$-B_{1, 2}(T_0)H^2$ a linear positive
component $A_{1, 2}(T_0)H$ of magnetoresistance appears in the
vicinity of $T_N$ in the AF phase of Ho$_{0.5}$Lu$_{0.5}$B$_{12}$,
and coefficients $A$ and $B$ change jump-wise in moderate magnetic
field 15-25 kOe between ($A_1$, $B_1$) and ($A_2$, $B_2$) values during
the AF$_1$$-$AF$_2$ phase transition observed at $T_C$. Following the
arguments presented previously in \cite{41, 52, 53}, the appearance of
a linear positive contribution to MR in the antiferromagnetic
phase should be attributed to scattering of charge carriers on
spin density waves (SDW). In particular, in case of metallic
chromium which is the most known $3D$ itinerant antiferromagnet
with SDW (having an incommensurate magnetic structure), the
amplitude of linear positive magnetoresistance reaches 180$\%$ at
magnetic field of $H$=12 kOe \cite{59}. Similar effects have
been found recently \cite{47} in the magnetoresistance of
Tm$_{1-x}$Yb$_x$B$_{12}$ antiferromagnets with the same modulated
incommensurate magnetic structure ( $q$$=$$(1/2\pm\delta,
1/2\pm\delta, 1/2\pm\delta)$ with $\delta$=0.035) as in HoB$_{12}$
\cite{58, 60}. Thus, according to our analysis, the AF$_1$$-$AF$_2$
transition observed in Ho$_{0.5}$Lu$_{0.5}$B$_{12}$ at $T_C$ (fig.\hyperref[FigX16]{16c})
 may be considered as a modification of the spin-density-wave
structure, which manifests itself both in (\textit{i}) changes of
the charge carrier scattering on magnetic nanodomains consisting
from interconnected Ho$^{3+}$ ions (expressed by the negative
quadratic Langevin type MR component) and (\textit{ii}) through
the increase of the SDW amplitude with increasing magnetic field,
resulting to enhancement of charge carrier scattering on SDW
(expressed through the linear positive MR term). We note that the
stabilization and enhancement of SDW in external magnetic field is
predicted previously \cite{61, 62}. However, at the same
time, to our best knowledge, no theoretical description of charge
transport in the presence of an external magnetic field in
itinerant magnets with incommensurate SDW structure is available
to date, which restricts the possibility of a more detailed
microscopic analysis of the positive magnetoresistance effect in
the rare earth dodecaborides.

To summarize the results of this section, we want to discuss below
shortly the mechanisms responsible for the emergence of various
magnetic phases in the AF state of Ho$_x$Lu$_{1-x}$B$_{12}$
compounds having a simple face centered cubic crystal structure
(fig.\hyperref[FigX1]{1a}-\hyperref[FigX1]{c}). To explain the nature of intermediate phases in the AF
state of HoB$_{12}$, authors of \cite{46} proposed a model that
considers frustration effects in the \textit{fcc} lattice of
RB$_{12}$. However, when taking into account (\textit{i}) the
loosely bounded state of rare earth ions in the dodecaboride
matrix, (\textit{ii}) the transition into the cage-glass state of
RB$_{12}$ at liquid nitrogen temperatures \cite{39}, and
(\textit{iii}) the appearance of disorder in the arrangement of
rare earth ions (random off-site location of Ho$^{3+}$-ions inside
the $B_{24}$ truncated cubooctahedron) resulting to formation of
magnetic nanosize clusters in studied compounds, it becomes
possible to explain the numerous phase transformations in the AF
state as a function of temperature and external magnetic field.
Indeed, positional disorder in the arrangement of Ho$^{3+}$ ions
in $B_{24}$ truncated cubooctahedrons leads to a significant
dispersion of exchange constants (through indirect exchange, RKKY
mechanism). Strong local 4\textit{f}$-$\textit{5d} spin
fluctuations then cause the appearance of an extra factor $-$ the
polarization of \textit{5d} conduction band states (the
spin-polaron effect). Moreover, the transition from paramagnetic
to AF phase is accompanied by the appearance of induced spin
polarization (formation of ferrons, according to the terminology
used in \cite{63, 64}) and by stabilization of these SDW antinodes
in the RB$_{12}$ matrix. The spin-polarized \textit{5d}-component
of the magnetic structure (ferrons) is from one side very
sensitive to external magnetic field, and, from another side, the
applied field suppresses 4\textit{f}$-$\textit{5d} spin
fluctuations by destroying the spin-flip scattering process. Thus,
the complex \textit{H-T} phase diagram of Ho$_x$Lu$_{1-x}$B$_{12}$ magnets
may be explained in terms of the formation of a combined
magnetically ordered state of localized $4f$ moments of
Ho$^{3+}$-ions in combination with spin polarized local areas of
the \textit{5d} states $-$ ferrons involved in the formation of a
spin density waves. The presence of the spin polarization was
confirmed for HoB$_{12}$ in \cite{46} where ferromagnetic
component of the order parameter was found in the magnetic neutron
diffraction patterns. Moreover, even harmonics and hysteresis of
the Hall resistance were detected in the range 20 kOe$<$$H$$<$60 kOe
for HoB$_{12}$ and it was attributed to charge carriers scattering
on SDW \cite{48}.

\section*{V. Conclusions}

We have studied in detail the transverse magnetoresistance of the
model metallic system with loosely bound state of rare earth
magnetic ions (Ho$^{3+}$) embedded in large size cavities
($B_{24}$ truncated cubooctahedrons) of the boron sublattice of
Ho$_x$Lu$_{1-x}$B$_{12}$ substitutional solid solutions. It was
shown that positive as well as negative magnetoresistance can be
observed in measurements of single crystals of these cage-glass
materials. The nMR component which appears in the paramagnetic
state of these magnetic metals, may be explained in terms of
charge carriers scattering on nanosize clusters of Ho ions with AF
exchange and short range AF order inside these domains. An
enhancement of the nMR effect is observed in concentrated Ho-based
dodecaborides in the vicinity of Neel temperature, and the Yosida
type relation $-\Delta\rho/\rho$$\sim$$M^2$ between magnetoresistance
and magnetization is found to provide an adequate description of
this term if a Langevin type behavior of magnetization is present.
Moreover, a reduction of effective values of Ho-ion magnetic
moments in the range 3-9$\mu_B$ was found to develop both with
temperature lowering and under increase of holmium content. It was
shown in the MR analysis that the positive quadratic term
$\Delta\rho/\rho_{(m+)}$$=$$\mu_D^2H^2$ dominates for all solid
solutions of Ho$_x$Lu$_{1-x}$B$_{12}$ at intermediate temperatures
20-120 K in strong magnetic fields, allowing to estimate the
exponential behavior of drift mobility of charge carriers
$\mu_D$$\sim$$T^{-\alpha}$ ($\alpha$=1.3$-$1.7). In the AF state an
additional positive linear MR contribution $\Delta\rho/\rho$$\sim$$
A(T)H$ was found and it was attributed to charge carriers
scattering on SDW in the incommensurate magnetic structure of
these unusual antiferromagnets. In accordance with magnetic field
induced modification of the AF state which has been observed
recently in neutron scattering studies of HoB$_{12}$ \cite{46} we
argue in favor of a SDW$_1$-SDW$_2$ magnetic phase transition in
Ho$_{0.5}$Lu$_{0.5}$B$_{12}$ in external fields of 10$-$25 kOe.
The presented comprehensive MR analysis allows to reconstruct the
\textit{H-T} magnetic phase diagram of
Ho$_{0.5}$Lu$_{0.5}$B$_{12}$, and provide arguments in favor of a
superposition of two components, the 4\textit{f} (based on
Ho$^{3+}$ localized moments) and the itinerant \textit{5d} (based
on SDW) parts, which form the complex magnetic structure of
Ho$_x$Lu$_{1-x}$B$_{12}$ antiferromagnets.



\section*{ACKNOWLEDGMENTS}

We would like to thank V.V. Moshchalkov, A.V. Kuznetsov, G.E.
Grechnev, J. Stankiewicz and K. Siemensmeyer for numerous helpful
discussions. This study was supported by the Branch of Physical
Sciences of the Russian Academy of Sciences within the program
"Strongly Correlated Electrons in Semiconductors, Metals,
Superconductors, and Magnetic Materials", Young scientists Grant
of RF President No. MK$-$6427.2014.2, and Slovak Scientific Grant
Agencies VEGA$-2/$0106$/$13, APVV$-$0132$-$11.


\newpage
\section*{SUPPLEMENTARY INFORMATION}\label{SI}
internet files:
\setcounter{figure}{0}
\renewcommand{\thefigure}{SI.\arabic{figure}}
\begin{figure*}[h]
 \begin{center}
\includegraphics[width = 7cm]{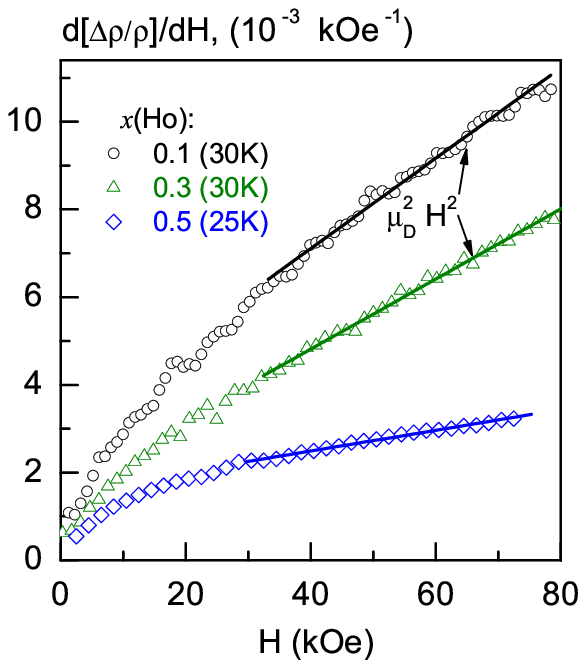}
    \end{center}
   \caption{The field dependences of derivatives of MR $d[\Delta\rho/\rho]/dH$ for solid solutions Ho$_x$Lu$_{1-x}$B$_{12}$ with $x$=0.1, 0.3 and 0.5. The positive quadratic term's approximation is illustrated by solid lines.}\label{FigXS1}
\end{figure*}

\begin{figure*}[h]
 \begin{center}
\includegraphics[width = 7cm]{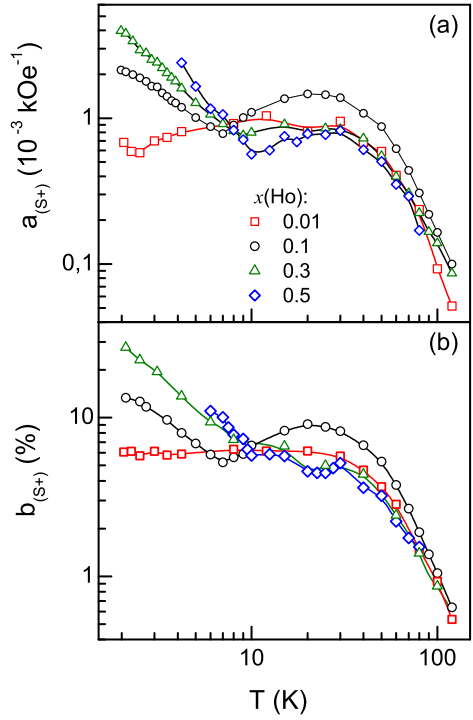}
   \caption{The temperature dependences of the amplitudes of (a) linear positive term $\sim$$a_{(S+)}$H and  (b) saturated component $b_{(S+)}$ of magnetoresistance for solid solutions Ho$_{x}$Lu$_{1-x}$B$_{12}$ with $x=$0.01, 0.1, 0.3 and 0.5 (see the text).}\label{FigXS2}
  \end{center}
\end{figure*}
\end{document}